\documentclass[twocolumn,superscriptaddress,showpacs,preprintnumbers,amsmath,amssymb]{revtex4}
\usepackage{amsmath}
\usepackage{amssymb}
\usepackage{dcolumn}
\usepackage{graphicx}
\usepackage{bm}
\usepackage{epstopdf}
\usepackage{threeparttable}
\usepackage{float}
\usepackage{color}
\usepackage{cleveref}

\usepackage{appendix}

\usepackage [english]{babel}
\usepackage [autostyle, english = american]{csquotes}
\MakeOuterQuote{"}
\usepackage[normalem]{ulem}

\makeatletter
\newcommand*{\rom}[1]{\expandafter\@slowromancap\romannumeral #1@}
\makeatother
\makeatletter
\def\p@subsection{}
\makeatother

\begin{document}
\title{Comparison of Shear and Compression Jammed Packings of Frictional Disks}

\author{Fansheng Xiong}
\affiliation{Zhou Pei-Yuan Center for Applied Mathematics, Tsinghua University, Beijing 100084, China}
\affiliation{Department of Mechanical Engineering and Materials Science, Yale University, New Haven, Connecticut, 06520, USA}
\author{Philip Wang} 
\affiliation{Department of Mechanical Engineering and Materials Science, Yale University, New Haven, Connecticut, 06520, USA}
\author{Abram H. Clark}
\affiliation{Department of Physics, Naval Postgraduate School, Monterey, California 93943, USA}
\author{Thibault Bertrand}
\affiliation{Department of Mathematics, Imperial College London, South Kensington Campus, London SW7 2AZ, England, UK}
\author{Nicholas T. Ouellette}
\affiliation{Department of Civil and Environmental Engineering, Stanford University, Stanford, California 94305, USA}
\author{Mark D. Shattuck}
\affiliation{Department of Physics and Benjamin Levich Institute, The City College of the City University of New York, New York, New York, 10031, USA}
\affiliation{Department of Mechanical Engineering and Materials Science, Yale University, New Haven, Connecticut, 06520, USA}
\author{Corey S. O'Hern}
\affiliation{Department of Mechanical Engineering and Materials Science, Yale University, New Haven, Connecticut, 06520, USA}
\affiliation{Center for Research on Interface Structures and Phenomena, Yale University, New Haven, Connecticut, 06520, USA}
\affiliation{Department of Physics, Yale University, New Haven, Connecticut, 06520, USA}
\affiliation{Department of Applied Physics, Yale University, New Haven, Connecticut, 06520, USA}

\date{\today}

\begin{abstract}
We compare the structural and mechanical properties of mechanically
stable (MS) packings of frictional disks in two spatial dimensions
(2D) generated with isotropic compression and simple shear protocols
from discrete element modeling (DEM) simulations.  We find that the
average contact number and packing fraction at jamming onset are
similar (with relative deviations $< 0.5\%$) for MS packings generated via
compression and shear. In contrast, the average stress anisotropy
$\langle {\hat \Sigma}_{xy} \rangle = 0$ for MS packings generated via
isotropic compression, whereas $\langle {\hat \Sigma}_{xy} \rangle >0$
for MS packings generated via simple shear. To investigate the
difference in the stress state of MS packings, we develop
packing-generation protocols to first unjam the MS packings, remove
the frictional contacts, and then rejam them.  Using these protocols,
we are able to obtain rejammed packings with nearly identical particle
positions and stress anisotropy distributions compared to the original
jammed packings. However, we find that when we directly compare the
original jammed packings and rejammed ones, there are finite
stress anisotropy deviations $\Delta {\hat \Sigma}_{xy}$. The
deviations are smaller than the stress anisotropy fluctuations
obtained by enumerating the force solutions within the null space of
the contact networks generated via the DEM simulations. These results
emphasize that even though the compression and shear jamming protocols
generate packings with the same contact networks, there can be
residual differences in the normal and tangential forces at each
contact, and thus differences in the stress anisotropy.

\end{abstract}

\maketitle

\section{Introduction} 
\label{sec:intro}
Granular materials, which are collections of macroscopic-sized grains,
can exist in fluidized states when the applied stress exceeds the
yield stress or in solid-like, or jammed, states when the applied
stress is below the yield stress ~\cite{Ref1,Ref2}. Many recent studies ~\cite{Ref3,Ref4,Ref5,Ref6,Ref7,Ref8} have shown that
the structural and mechanical properties of jammed granular packings
depend on the protocol that was used to generate them.  For example,
when granular packings are generated via simple or pure shear, the
force chain networks appear more heterogeneous and anisotropic.  In
contrast, for granular packings generated via isotropic compression,
the force distribution is more uniform ~\cite{Ref9,Ref10,Ref11,Ref12,Ref13}.  This protocol dependence
for the structural and mechanical properties of jammed packings makes
it difficult to acccurately calculate, and even properly define, their
statistical averages.

An important question to address when considering how to calculate
statistical averages of a system's structural and mechanical
properties is to determine which states are to be included in the
statistical ensemble. For jammed granular packings, the relevant set
of states is the collection of mechanically stable (MS) packings
~\cite{Ref14,Ref15} with force and torque balance on every grain. In
addition, the average properties of the ensemble of MS packings depend 
on the probabilities with which each MS packing occurs, and the 
probabilities can vary strongly with the packing-generation protocol.

We recently investigated how the mechanical properties of granular
systems composed of bidisperse frictionless disks interacting via
pairwise, purely repulsive central forces ~\cite{Ref16} depend on the
packing-generation protocol. In this case, the relevant ensemble of
jammed states is the collection of isostatic MS packings
~\cite{Ref16,Ref17,Ref18,Ref19} with $N_c = 2N'-1$ interparticle
contacts, where $N'=N-N_r$, $N$ is the number of disks, and $N_r$ is
the number of rattler disks with less than $3$ contacts.  We compared
MS packings of frictionless disks generated via simple or pure shear
(i.e.  shear jammed packings) and those generated via isotropic
compression (i.e. compression jammed packings).  We found that
compression jammed packings can possess either positive or negative
stress anisotropy ${\hat \Sigma}_{xy} = -\Sigma_{xy}/P$, where
$\Sigma_{xy}$ is the shear stress and $P$ is the pressure of the MS
packing.  In contrast, shear jammed MS packings possess only ${\hat
  \Sigma}_{xy} >0$ and these packings are identical to the MS packings
generated via isotropic compression with ${\hat \Sigma}_{xy}
>0$. Thus, the ensemble of jammed packings generated via shear and
isotropic compression is the same, but shear (in one direction)
selects jammed packings with only one sign of the stress anisotropy.

In this article, we will investigate a similar question of whether
exploring configuration space through shear versus through compression
samples the same set of MS packings, except we consider the case of
jammed packings of dry, frictional disks. A key feature of frictional
systems is that the forces at each interparticle contact must obey the
Coulomb condition ~\cite{Ref20,Ref21}, where $f^t_{ij} \le \mu f^n_{ij}$,
$f^n_{ij}$ and $f^t_{ij}$ are the normal and tangential forces at the
contact between particles $i$ and $j$, and $\mu$ is the static
friction coefficient. If $f^t_{ij}$ exceeds $\mu f^n_{ij}$, the
contact will slide to satisfy the Coulomb condition. Further, the
number of contacts for MS packings of frictional disks is below the
isostatic value $z_{\rm iso}=4$, and thus there are many solutions for
the normal and tangential forces for each fixed network of
interparticle contacts. Thus, one can imagine that different protocols
for generating jammed packings of frictional disks can give rise to MS
packings with different distributions of sliding contacts, different
force solutions for a given contact network, or even different types of contact
networks.

We carry out discrete element modeling (DEM) simulations of bidipserse
frictional disks in two dimensions (2D) to compare the properties of
MS packings at jamming onset generated via simple shear and isotropic
compression. We find five significant results: 1) The average packing
fraction $\langle \phi_J(\mu) \rangle$ and contact number $\langle z_J
(\mu)\rangle$ at jamming onset versus friction coefficient $\mu$ for
the ensemble of MS packings generated via isotropic compression and
simple shear are similar (with deviations $< 0.5\%$). In particular,
both shear and compression jammed packings can possess a range of
average contact numbers $\langle z_J\rangle$ between $3$ and $4$,
depending on $\mu$. 2) As with frictionless disks, we find that MS
packings of frictional disks generated via isotropic compression
possess both ${\hat \Sigma}_{xy} > 0$ and ${\hat \Sigma}_{xy} <0$,
whereas MS packings generated via simple shear possess only one sign
of the stress anisotropy.  3) For each MS packing generated via simple
shear, we can decompress the packing to remove all of the frictional
contacts and recompress it to generate an MS packing with particle
positions that are nearly identical to those of the original shear
jammed MS packing.  Similarly, for each MS packing generated via
isotropic compression, we can shear it in a given direction to unjam
it and remove all of the frictional contacts and shear it back in the
opposite direction to generate an MS packing with disk positions that
are nearly identical to those of the original compression jammed
packing. 4) Even though the disk positions are nearly identical, we
find a small, but significant difference between the stress anisotropy
of the shear jammed packings and that for the compression rejammed
packings.  Similarly, we find a smaller, but significant difference in
the stress anisotropy between the compression jammed packings and that
for the shear rejammed packings. The fluctuations in the stress
anisotropy from the DEM simulations is smaller than the fluctuations
obtained by enumerating all normal and tangential forces solutions
from the null space for each fixed contact network. 5) We also show
that even though we can generate MS packings with nearly identical
particle positions via the DEM simulations, the packings can possess
very different mobility distributions $P(\xi)$, where $\xi =
F^t_{ij}/\mu F^n_{ij}$, and numbers of sliding contacts.

The remainder of the article is organized as follows. The Methods
section (Sec.~\ref{sec:1}) introduces the Cundall-Strack model
~\cite{Ref22} for static friction between disks, the definitions of
the stress tensor, shear stress, and stress anisotropy, and the
details of the isotropic compression and simple shear packing
generation protocols. In addition, we describe the protocols to
decompress and then recompress shear-jammed packings and shear unjam
and then shear jam compression-jammed packings.  The Results section
(Sec.~\ref{sec:2}) describes our findings for the average packing
fraction and contact number at jamming onset versus the static
friction coefficient for MS packings generated via both protocols.  In
addition, we show the stress anisotropy and mobility distributions for
each protocol that we use to generate MS packings. In the Conclusion
and Future Directions section (Sec.~\ref{sec:4}), we summarize our
results and describe promising future research directions, e.g. enuerating 
the force solutions for the null space of contact networks generated 
via isostropic compression and shear.

\begin{figure}[h]
\resizebox{1.0\hsize}{!}{%
  \includegraphics{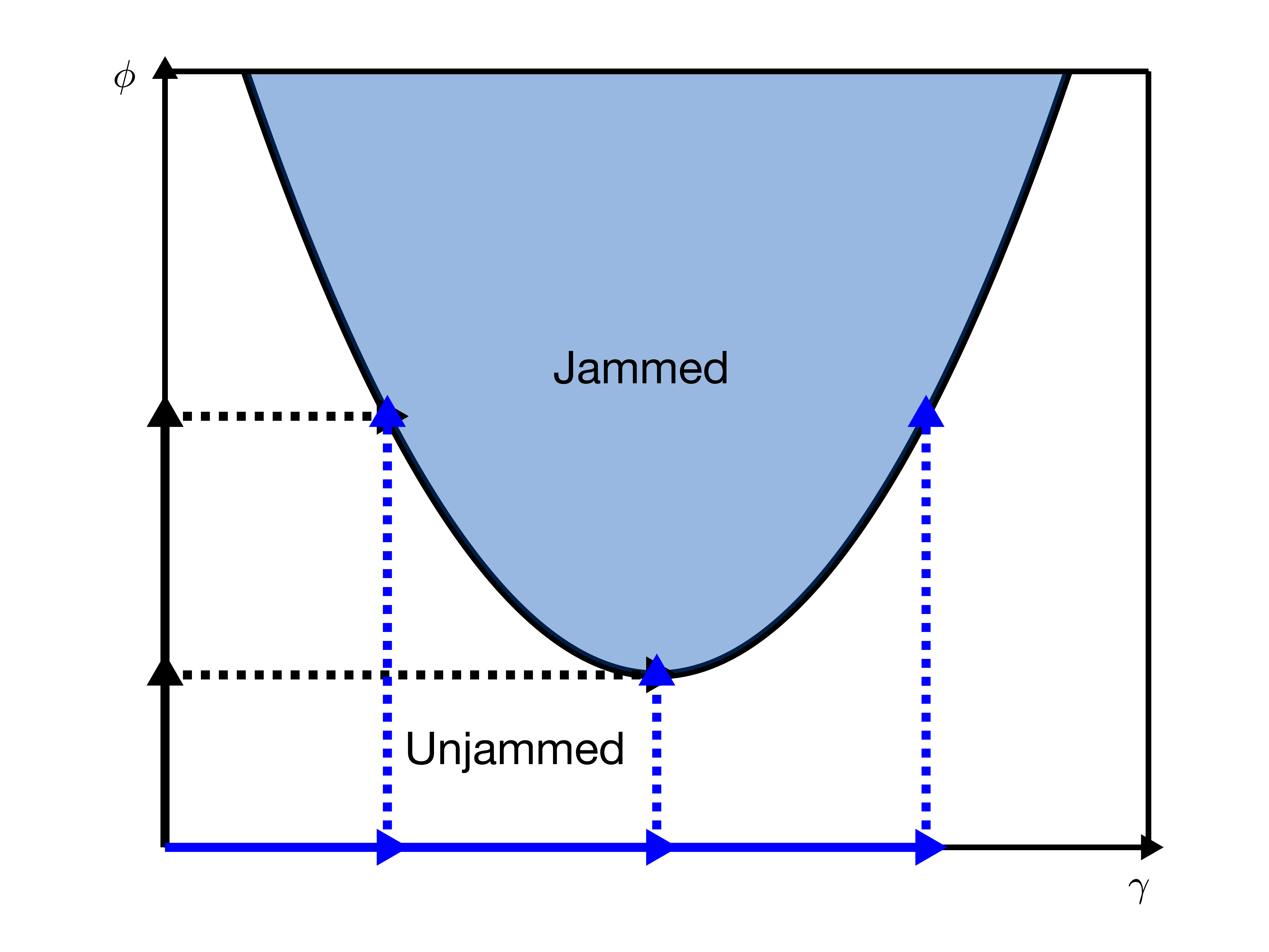}
}
\caption{An idealized jamming diagram in which the jammed and unjammed
regions are separated by a parabolic boundary in the packing
fraction $\phi$ and shear strain $\gamma$ plane. For compression
jamming, we first apply simple shear strain $\gamma$ at $\phi=0$
(horizontal solid blue lines) and then compress the system at fixed
$\gamma$ to jamming onset at $\phi_J$ (vertical dashed blue
lines). For shear jamming, we first compress the system to $\phi <
\phi_J$ (vertical solid black lines) and then apply simple shear to
jamming onset at $\gamma_J$ (horizontal dashed black lines).}
\label{fig:1}       
\end{figure}


\begin{figure}[h]
\resizebox{1.0\hsize}{!}{%
  \includegraphics{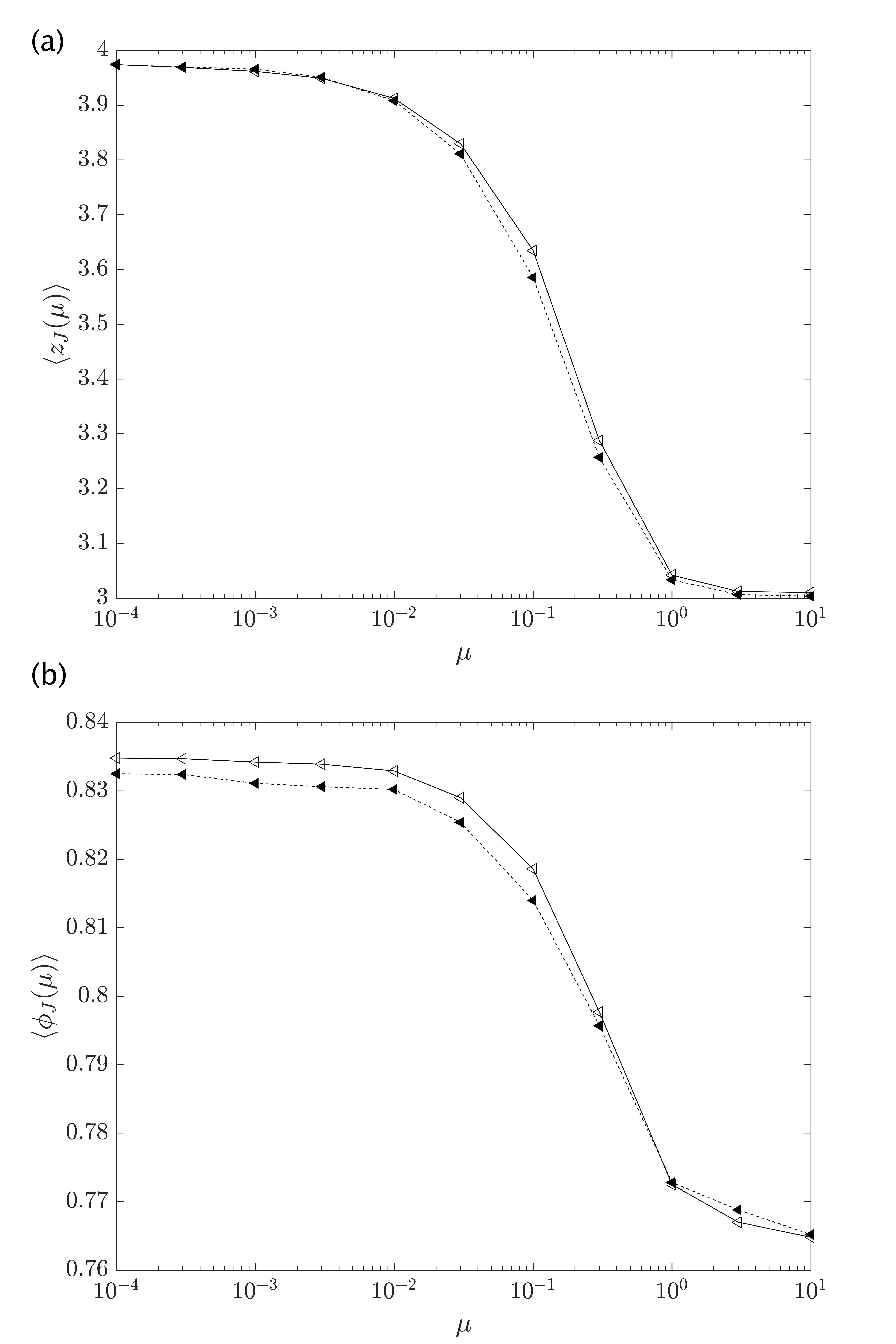}
}
\caption{Average (a) contact number $\langle z_J \rangle$ and (b) 
packing fraction $\langle \phi_J \rangle$ at jamming onset for MS packings 
generated via simple shear (filled triangles; dotted lines) and isotropic 
compression (open triangles; solid lines) plotted versus the static 
friction coefficient $\mu$ for $N=128$ bidisperse frictional disks. The averages were calculated 
over more than $50$ independent MS packings at each $\mu$.}
\label{fig:2}       
\end{figure}



\section{methods} 
\label{sec:1}
We perform DEM simulations of frictional disks in 2D. We consider
bidisperse mixtures of disks with $N/2$ large disks and $N/2$ small
disks, each with the same mass $m$, and diameter ratio
$\sigma_{l}/\sigma_{s} = 1.4$ ~\cite{Ref23}.  The MS packings are generated inside a
square box with side length $L$ and periodic boundary conditions in
both directions. The disks interact via pair forces in the normal
(along the vector ${\hat r}_{ij}$ from the center of disk $j$ to that
of disk $i$) and the tangential ${\hat t}_{ij}$ directions (with
${\hat t}_{ij} \cdot {\hat r}_{ij} =0$).  We employ a repulsive linear
spring potential for forces in the
normal direction:
\begin{equation}
U^n(r_{ij}) = \frac{K \sigma_{ij}}{2} \left(1 - \frac{r_{ij}}{\sigma_{ij}}
\right)^2 \theta\left( 1 - \frac{r_{ij}}{\sigma_{ij}} \right),
\end{equation}
where $r_{ij}$ is the separation between disk centers, $\sigma_{ij} =
(\sigma_i +\sigma_j)/2$, $\sigma_i$ is the diameter of disk $i$, $K$
is the spring constant in the normal direction, and $\theta(.)$ is the
Heaviside step function that sets the interaction potential to zero
when disks $i$ and $j$ are not in contact.

We implement the Cundall-Strack model ~\cite{Ref22} for the tangential frictional
forces.  When disks $i$ and $j$ are in contact, ${\vec f}^t_{ij} = K_t
{\vec u}^t_{ij}$, where $K_t = K/3$ is the spring constant for
the tangential forces and ${\vec u}^t_{ij}$ is the relative tangential
displacement.  ${\vec u}^t_{ij}$ is obtained by inegrating the relative
tangential velocity~\cite{Ref24,Ref25}, while disks $i$ and $j$ are in contact:
\begin{equation}
 \frac{d{\vec u}_{ij}^{t}}{dt}={\vec v}_{ij}^{t}-\frac{({\vec u}_{ij}^{t}\cdot{\vec v}_{ij}){\vec r}_{ij}}{r_{ij}^{2}},
\end{equation}
where ${\vec v}_{ij} = {\vec v}_i-{\vec v}_j$, ${\vec
  v}_{ij}^{t}={\vec v}_{ij} - {\vec v}^n_{ij} -\frac{1}{2}({\vec
  \omega}_i +{\vec \omega}_j)\times {\vec r}_{ij}$, ${\vec v}^n_{ij} =
({\vec v}_{ij} \cdot {\hat r}_{ij}) {\hat r}_{ij}$, and ${\vec
  \omega}_i$ is the angular velocity of disk $i$.  ${\vec u}_{ij}^{t}$ is
set to zero when the pair of disks $i$ and $j$ is no longer in
contact.  We implement the Coulomb criterion, $f_{ij}^{t} \leq \mu
f_{ij}^{n}$, by resetting $|{\vec u}^t_{ij}| = u^t_{ij} = \mu f^n_{ij}/K_t$ if
$f_{ij}^{t}$ exceeds $\mu f_{ij}^{n}$.  The total potential energy is
$U=U^n+U^t$, where $U^n = \sum_{i>j} U^n(r_{ij})$ and $U^t =
\sum_{i>j} K_t (u^t_{ij})^2$.

We characterize the stress of the MS packings using the virial
expression for the stress tensor~\cite{Ref16}:
\begin{equation}
\label{stress_tensor}
\Sigma_{\beta \delta} = \frac{1}{A} \sum_{i > j}f_{i j \beta}r_{i j \delta},
\end{equation}
where $\beta$, $\delta=x$, $y$, $A=L^2$ is the area of the simulation
box, $f_{i j \beta}$ is the $\beta$-component of the interparticle
force ${\vec f}_{ij}$ on disk $i$ due to disk $j$, and $r_{ij\delta}$ is the
$\delta$-component of the separation vector ${\vec r}_{ij}$. We define
the stress anisotropy as ${\hat \Sigma}_{xy} = -\Sigma_{xy}/P$ and the pressure as
$P=(\Sigma_{xx}+\Sigma_{yy})/2$.  We measure length, energy, and stress 
below in units of $\sigma_s$, $K \sigma_s$, and $K/\sigma_s$, respectively. 

We employ two main protocols to generate MS packings: 1) isotropic
compression at fixed shear strain $\gamma$ and 2) simple shear at
fixed packing fraction $\phi$.  (See Fig.~\ref{fig:1}.) For protocol
$1$ (isotropic compression), we first randomly place the disks in the
simulation cell without overlaps. We then increase the diameters of
the disks according to $\sigma_i'=\sigma_i(1+d\phi/\phi)$ where
$d\phi<10^{-4}$ is the initial increment in the packing fraction.
After each small change in packing fraction, we minimize the total potential
energy $U$ by adding viscous damping forces proportional to each
disk's velocity ${\vec v}_i$.  Energy minimization is terminated when
$K_{\rm max}<10^{-20}$, where $K_{\rm max}$ is the maximum kinetic
energy of one of the disks.

If $U/N < U_{\rm tol}$ after minimization, we increase the packing
fraction again by $d\phi$ and then minimize the total potential
energy. To eliminate overlaps, we typically set $U_{\rm tol}
=10^{-16}$, which means that the typical disk overlap is $<
10^{-8}$. If after minimization, $U/N>2U_{\rm tol}$, the growth step
is too large and we return to the uncompressed packing of the previous
step with $U/N < U_{\rm tol}$.  Instead, we increase the packing
fraction by $d\phi/2$, and minimize the total potential energy. We
repeat this process until the total potential energy satisfies $U_{\rm
  tol}<U/N<2U_{\rm tol}$, at which we assume that the packing has
reached jamming onset at packing fraction $\phi_J$. This compression
protocol ensures that the system approaches jamming onset from below.

For protocol $2$, we first prepare the system below jamming onset at
$\phi_t < \phi_J$ (using protocol $1$). We then apply successive
simple shear strain increments $d\gamma$ by shifting the disk
positions, $x_i' = x_i + d\gamma y_i$, and implementing Lees-Edwards
boundary conditions, which are consistent with the applied affine
shear strain. The initial shear strain increment is $d\gamma=10^{-4}$.
After an applied shear strain increment, we minimize the total
potential energy. Energy minimization is again
terminated when $K_{\rm max}<10^{-20}$. If $U/N < U_{\rm tol}$ after
minimization, we increment the shear strain again by $d\gamma$ and
minimize the total potential energy. If after minimization,
$U/N>2U_{\rm tol}$, the shear strain step is too large and we return
to the packing at the previous strain step with $U/N < U_{\rm tol}$.
Instead, we increment the shear strain by $d\gamma/2$, and minimize
the total potential energy. We repeat this process until the total
potential energy satisfies $U_{\rm tol}<U/N<2U_{\rm tol}$, at which we
assume that the packing has reached jamming onset at total shear
strain $\gamma_J$.

Energy minimization is carried out by integrating Newton's equations 
of motion for the translational and rotational degrees of freedom of each 
disk in the presence of static friction and viscous dissipation. For 
the translational degrees of freedom, we have  
\begin{equation}
\label{newton}
m \frac{d^2{\vec r}_i}{dt^2} = {\vec f}_i^n+{\vec f}^t_i + {\vec f}^d_i, 
\end{equation}
where ${\vec f}^n_i = \sum_j {\vec f}^n_{ij}$, ${\vec f}^n_{ij} =
-dU^n/d{\vec r}_{ij}$, ${\vec f}^t_i = \sum_j {\vec f}^t_{ij}$, ${\vec
  f}^d_i = -b^n {\vec v}_i$, $b^n$ is the damping coefficient, and the
sums over $j$ include disks that are in contact with disk $i$. For
the rotational degrees of freedom, we have
\begin{equation}
\label{rotation}
I_i \frac{d {\vec \omega}_i}{dt} = {\vec \tau}_i - b^t {\vec \omega}_i,  
\end{equation}
where $I_i = m\sigma_i^2/8$ is the moment of inertia for disk $i$, 
$b^t$ is the rotational damping coefficient, and   
\begin{equation}
\label{torque}
{\vec \tau}_i = \frac{1}{2} \sum_j {\vec r}_{ij} \times {\vec F}^t_{ij}
\end{equation} 
is the torque on disk $i$. We chose $b^n$ and $b^t$ so that the
dynamics for the translational and rotational degrees of freedom are
in the overdamped limit.

After generating MS packings using these two protcols, we measure the
contact number $z=N_c/N'$, where $N_c$ is the total number of contacts
in the system, and shear stress anisotropy of the MS packings. For
these measurements, we recursively remove rattler disks with fewer
than three contacts for frictionless disks or fewer than two contacts
for frictional disks.

\section{Results} 
\label{sec:2}

In this section, we first describe our results for the average contact
number and packing fraction of MS packings generated via isotropic
compression and simple shear.  We then explain why the distribution of
the shear stress anisotropy differs for compression and shear jammed
packings. We also develop a protocol where we unjam shear jammed
packings and then re-jam them via isotropic compression and a protocol
where we unjam compression jammed packings and then re-jam them via
applied shear strain.  We then compare the contact network and stress
anisotropy of the original jammed packings and the re-jammed packings,
and show that the disk positions of the re-jammed packings are nearly
identical to those for the original jammed packings.  We find small
differences in the stress state of the original jammed packings and
the rejammed ones, but these differences are smaller than the
fluctuations obtained by enumerating all of the normal and tangential
force solutions for a given jammed packing consistent with force and
torque balance.

\subsection{Packing fraction and contact number}
\label{sec:2.1}

In Fig.~\ref{fig:2}, we show (for $N=128$) that the contact number
$\langle z_J \rangle$ and packing fraction $\langle \phi_J \rangle$ at
jamming onset are similar for compression and shear jammed packings
over the full range of friction coefficients $\mu$. (The relative 
deivations are less than $0.5\%$.) For both
protocols, we find that $z \approx 4$ in the small-$\mu$ limit and
$z\approx 3$ in the large-$\mu$ limit, as found previously in several
numerical studies of frictional disks.  The average packing fraction
$\langle \phi_J \rangle \approx 0.835$ in the small-$\mu$ limit and
$\approx 0.765$ in the large-$\mu$ limit. The crossover between the
low- and high-friction behavior in the contact number and packing
fraction again occurs near $\mu_c \approx 0.1$ for both protocols. 
This crossover value of $\mu$ is similar to that found previously in 
compression jammed frictional disk packings~\cite{Ref17}.

The average packing fraction at jamming onset is slightly smaller for
shear jammed packings compared to that for compression jammed
packings.  This small difference in packing fraction stems from
differences in the compression and shear jamming protocols.  For each
initial condition $i$, we generate a compression jammed packing with
$\phi_J^{i}$. Then, for each $i$, we generate a series of unjammed 
configurations
with $\phi^i_{\alpha} < \phi_J^{i}$ and shear them until they jam at
$\gamma_J$.  To obtain $\langle \phi_J \rangle$ for the shear jamming
protocol, we average $\phi^i_{\alpha}$ over $i$ and $\alpha$ for all
systems that jammed. This protocol for generating shear jammed
packings is thus biased towards finding MS packings with packing
fractions lower than those found for isotropic compression.  Despite
this, the packing fraction at jamming onset $\langle \phi_J(\mu)
\rangle$ for the two protocols differs by less than $0.5\%$ over the 
full range of $\mu$.

\begin{figure}
\resizebox{1.0\hsize}{!}{%
  \includegraphics{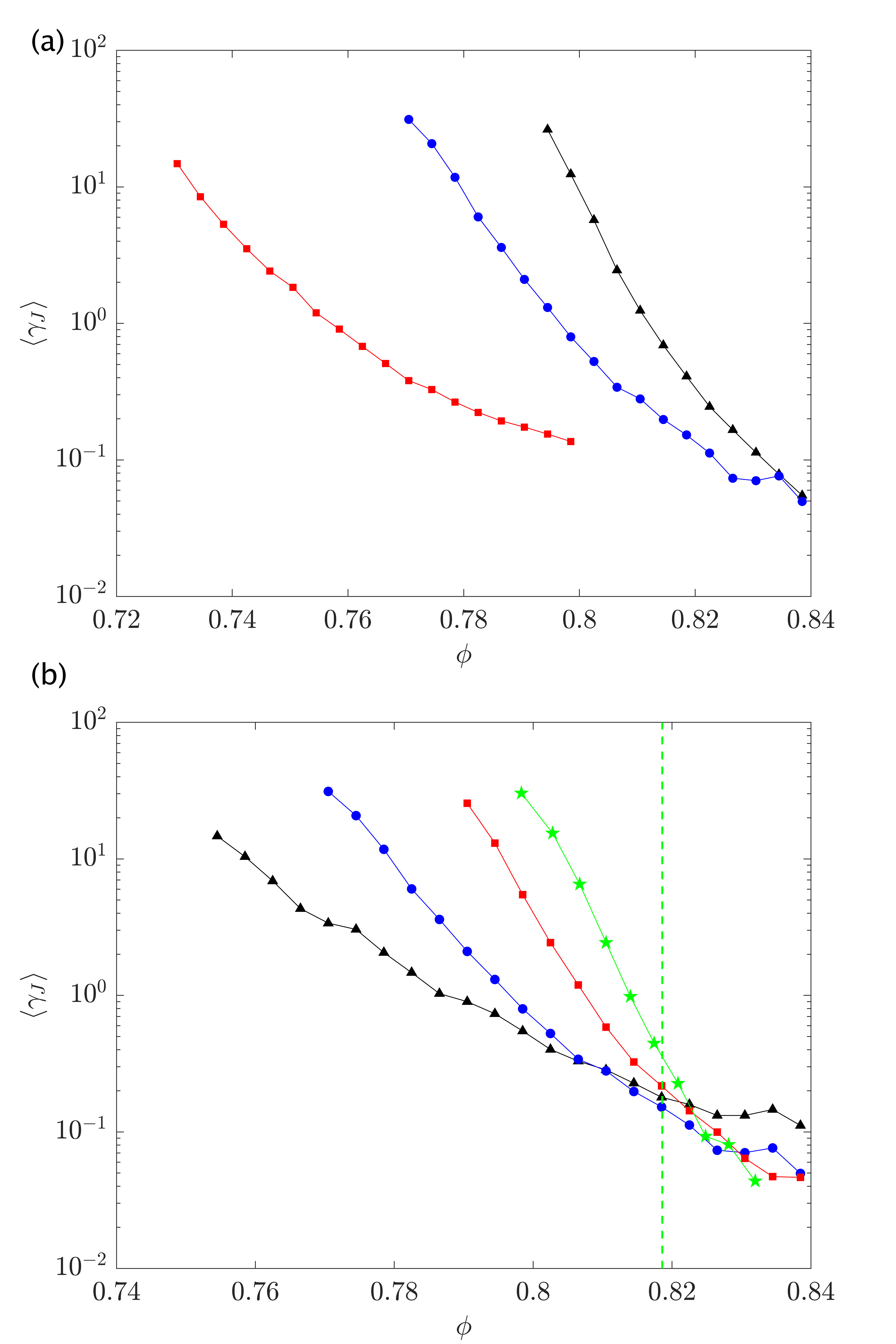}
}
\caption{Average total shear strain $\langle \gamma_J\rangle$ required 
to jam a collection of disks with (a) $N=32$ as a function of 
packing fraction $\phi$ for several  
friction coefficients, $\mu=0$ (black triangles), $0.1$ (blue circles), and $1.0$ 
(red squares) and for (b) $\mu=0.1$ and several  
system sizes, $N=16$ (black triangles), $32$ (blue circles), 
$64$ (red squares), and $128$ (green stars). The vertical dashed line 
indicates $\langle 
\phi_J \rangle$ for compression jammed packings with $\mu = 0.1$ and $N=64$.}
\label{fig:3}       
\end{figure}

In Fig.~\ref{fig:3}, we show the average shear strain $\langle
\gamma_{J} \rangle$ required to find a jammed packing starting from an
initially unjammed packing using the shear jamming protocol as a
function of packing fraction. In panel (a), we plot $\langle \gamma_J
\rangle$ versus $\phi$ for several friction
coefficients.  The average strain increases with decreasing packing
fraction and the range of packing fractions over which a shear jammed
packing can be obtained shifts to lower values with increasing
friction coefficient. In panel (b), we show $\langle \gamma_J \rangle$
versus $\phi$ at $\mu =0.1$ and several system
sizes.  We find that the slope $d\langle \gamma_J \rangle/d\langle
\phi_J \rangle$ increases with increasing system size.  For the
$\mu=0.1$ data in panel (b), we expect $\langle \gamma_J\rangle$ to
become vertical near $\phi \approx 0.82$, which is $\langle
\phi_J(\mu) \rangle$ for compression jammed packings, in the
large-system limit. The system-size dependence of $\langle \gamma_J
\rangle$ is similar to that found for packings of frictionless disks.
Thus, we predict that the range of packing fraction over which shear
jamming occurs to shrink with increasing system size. In particular, we 
expect shear jamming to occur over a narrow range of packing fraction 
near $\langle \phi_J(\mu) \rangle$ obtained from isotropic compression in 
the large-system limit. 

%

\subsection{Stress anisotropy of compression and shear jammed packings}
\label{sec:2.2}

\begin{figure}
\resizebox{1.0\hsize}{!}{%
  \includegraphics{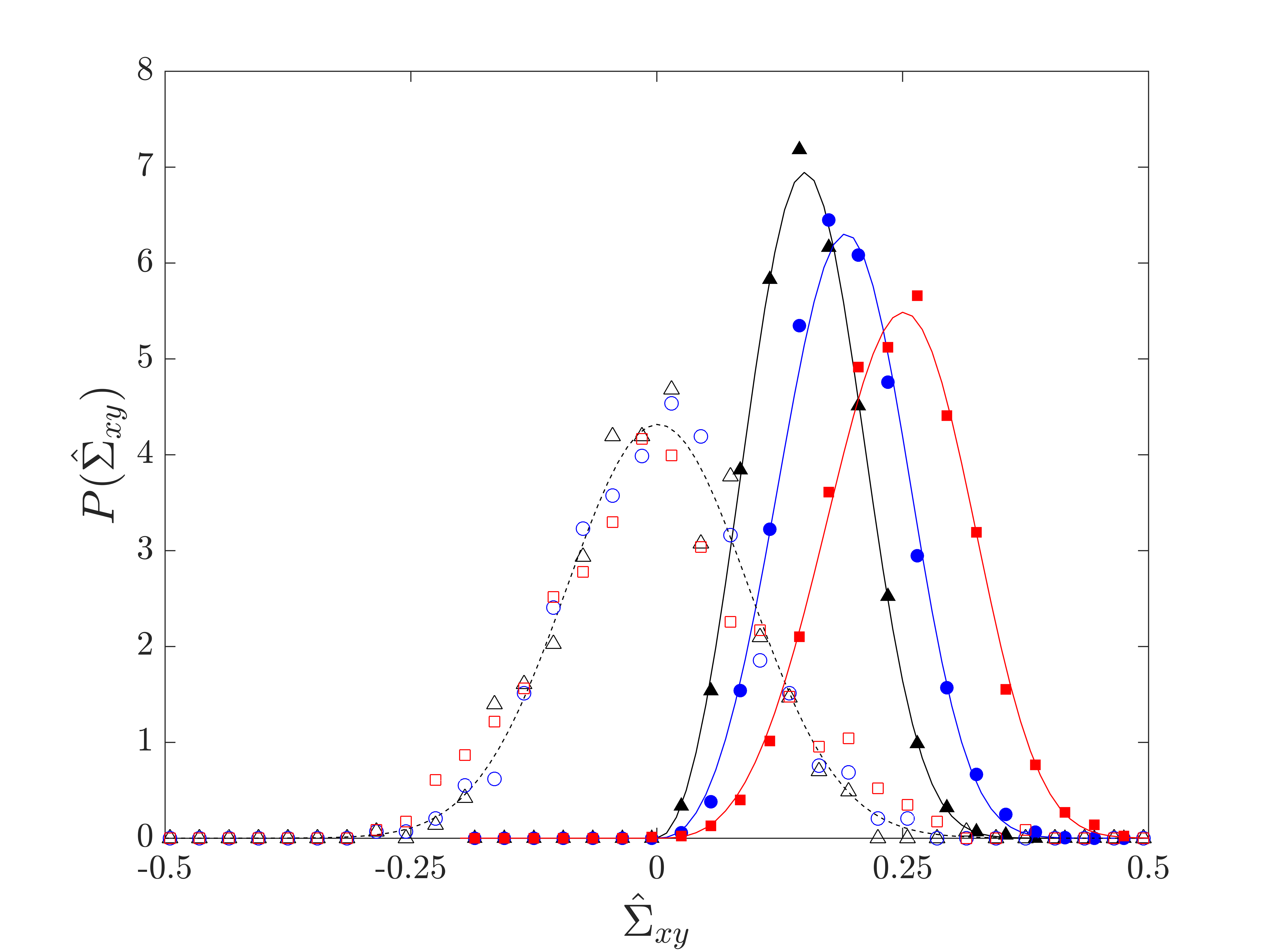}
}
\caption{Probability distributions of the shear stress anisotropy 
${\hat \Sigma}_{xy}$ for packings generated via isotropic compression 
(open symbols) and simple shear (filled symbols).
For both packing-generation protocols, we show distributions for $N=64$ and 
friction coefficients $\mu=0$ (triangles), $0.1$ 
(circles), and $1.0$ (squares). The distributions 
were obtained from more than $10^{3}$ independently generated jammed packings.
The dashed line is a Gaussian distribution with zero mean and standard 
deviation $\Delta \sim 0.1$ and the solid lines are Weibull distributions 
with scale and shape parameters $\lambda \sim 0.17$ and $k\sim 3.0$, 
$\lambda \sim 0.21$ and $k \sim 3.5$, and $\lambda \sim 0.27$ and $k \sim 3.9$ from left to right.}
\label{fig:4}       
\end{figure}

\begin{figure}[h]
\resizebox{1.0\hsize}{!}{%
  \includegraphics{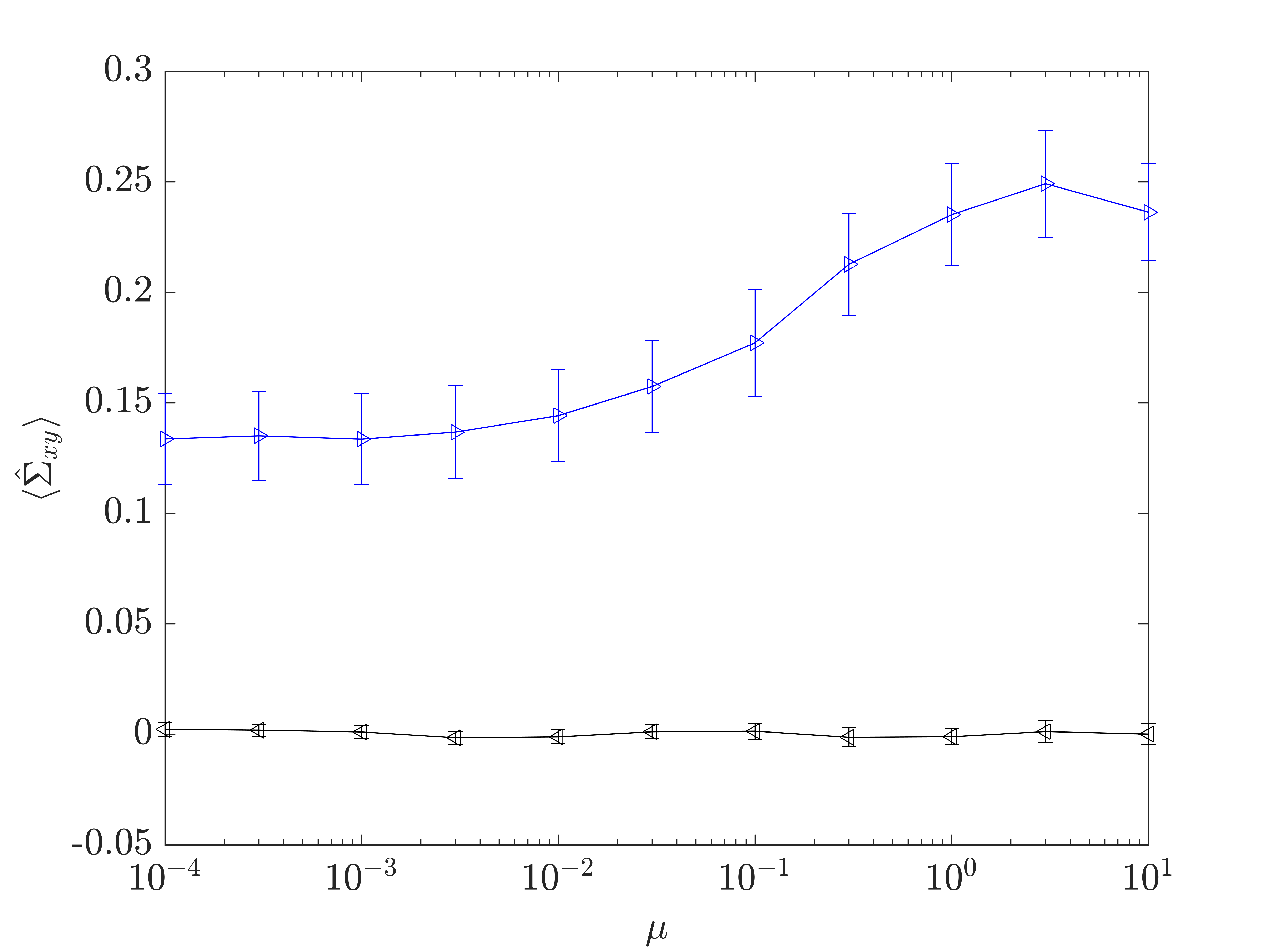}
}
\caption{Average shear stress anisotropy $\langle \hat{\Sigma}_{xy} \rangle$ 
at jamming onset for MS packings generated via simple shear 
(filled triangles) and isotropic compression (open triangles) plotted 
versus the static friction coefficient $\mu$ for $N=128$. The error bars
indicate the standard deviation in $P({\hat \Sigma}_{xy})$ for 
each protocol.}
\label{fig:5}       
\end{figure}

In previous studies, we showed that a significant difference between
shear and compression jammed packings of frictionless disks is that
shear jammed packings possess a non-zero average shear stress
anisotropy $\langle {\hat \Sigma}_{xy} \rangle >0$, whereas
compression jammed packings possess $\langle {\hat \Sigma}_{xy}\rangle
=0$. We find similar behavior for MS packings of frictional disks. In
Fig.~\ref{fig:4}, we show the distribution of shear stress anisotropy
$P({\hat \Sigma}_{xy})$ for packings with three friction coefficients
$\mu = 0$, $0.1$, and $1.0$ using the isotropic compression and shear
jamming protocols.  For the isotropic compression protocol, $P({\hat
  \Sigma}_{xy})$ is a Gaussian distribution with zero mean, whereas
${\hat \Sigma}_{xy} >0$ for packings generated via simple shear (in a
single direction). The stress anisotropy distributions $P({\hat
  \Sigma}_{xy})$ for simple shear are Weibull distributions with shape
and scale factors that depend on $\mu$~\cite{Ref26}. In Fig.~\ref{fig:5}, we show
the corresponding averages of the shear stress anisotropy
distributions. We find that $\langle {\hat \Sigma}_{xy} \rangle =0$
for all $\mu$ for packings generated using isotropic compression. In
contrast, for packings generated via simple shear, $\langle {\hat
  \Sigma}_{xy} \rangle \approx 0.13$~\cite{Ref27} for $\mu \rightarrow 0$ and
$\langle {\hat \Sigma}_{xy} \rangle$ increases with $\mu$ until
reaching $\langle {\hat \Sigma}_{xy} \rangle \approx 0.25$ in the
large-$\mu$ limit.
 
We also showed in previous studies~\cite{Ref15} that MS packings of frictionless
disks occur in geometrical families in the packing fraction $\phi$ and
shear strain $\gamma$ plane. For frictionless disks, geometrical
families are defined as MS packings with the same network of
interparticle contacts, with different, but related fabric tensors.
The packing fractions of MS packings in the same geometrical family
are related via $\phi = \phi_0 + A (\gamma-\gamma_0)^2$, where $A>0$
is the curvature in the $\phi$-$\gamma$ plane, and $\phi_0$ is the
minimum value of the packing fraction at strain $\gamma=\gamma_0$~\cite{Ref16}. The
parameters $A$, $\phi_0$, and $\gamma_0$ vary from one geometrical
family to another.

Using a general work-energy relationship for packings undergoing
isotropic compression and simple shear, we showed~\cite{Ref19} that for
packings of frictionless disks, the shear stress stress anisotropy 
can be obtained from the dilatancy, $d\phi_J/d\gamma$:
\begin{equation}
\label{slope}
{\hat \Sigma}_{xy} = -\frac{1}{\phi}\frac{d\phi_J}{d\gamma}.
\end{equation}
The isotropic compression protocol can sample packings with
alternating signs of $d\phi_J/d\gamma$ (and thus ${\hat \Sigma}_{xy} >
0$ and $<0$), whereas the shear jamming protocol can only sample
packings with $d\phi_J/ d\gamma < 0$ (and thus ${\hat \Sigma}_{xy} >
0$).  We expect similar behavior for packings of frictional disks,
however, it is more difficult to identify single geometrical
famailies. First, Eq.~\ref{slope} does not account for sliding
contacts, and thus geometrical families must be defined over
sufficiently small strain intervals such that interparticle contacts
do not slide.  In addition, for each MS packing of frictional disks in
a given geometrical family, there is an ensemble of solutions for the
normal and tangential forces~\cite{Ref20}, not a unique solution, as for the normal
forces in packings of frictionless disks. The extent to which packings
with the same contact networks (and particle positions) can possess
different shear stress anisotropies will be discussed in more detail
in Sec.~\ref{sec:2.3} below.

\subsection{Unjam and rejam compression and shear jammed packings}\sloppy
\label{sec:2.3}

In Sec.~\ref{sec:2.1}, we showed that compression and shear jammed
packings have similar contact number $\langle z_J(\mu)\rangle$ and
packing fraction $\langle \phi_J(\mu)\rangle$ over the full range of
$\mu$. However, in Sec.~\ref{sec:2.2}, we demonstrated that $\langle {\hat
  \Sigma}_{xy} \rangle = 0$ for compression jammed packings and
$\langle {\hat \Sigma}_{xy} \rangle > 0$ for shear jammed packings.
Does this significant difference in the stress state of MS packings
occur because the packings generated via isotropic compression are
fundamentally different from those generated via simple shear?

To address this question, we consider two new protocols---protocol
$A$, where we decompress each shear jammed packing, releasing all of the
frictional contacts, and then re-compress each one until each jams,
and protocol $B$, where we shear unjam each compression jammed
packing, releasing all of the frictional contacts, and then shear each
one until each jams.  The goal is to study protocols that allow the
system to move away from a given jammed packing in configuration
space, removing all of the frictional contacts, and determine to what
extent the system can recover the original jammed packing using either
compression or shear.  We compare the particle positions, shear stress
anisotropy, and contact mobility for the original and re-jammed
packings. If there is no difference between the original jammed 
and re-jammed packings, all MS packings can be generated via compression 
or shear. For protocols $A$ and $B$, we will focus on systems with
$N=16$ and $\mu=0.1$, but we find similar results for systems with
larger $N$ and different $\mu$.

\begin{figure}
\resizebox{1.0\hsize}{!}{%
  \includegraphics{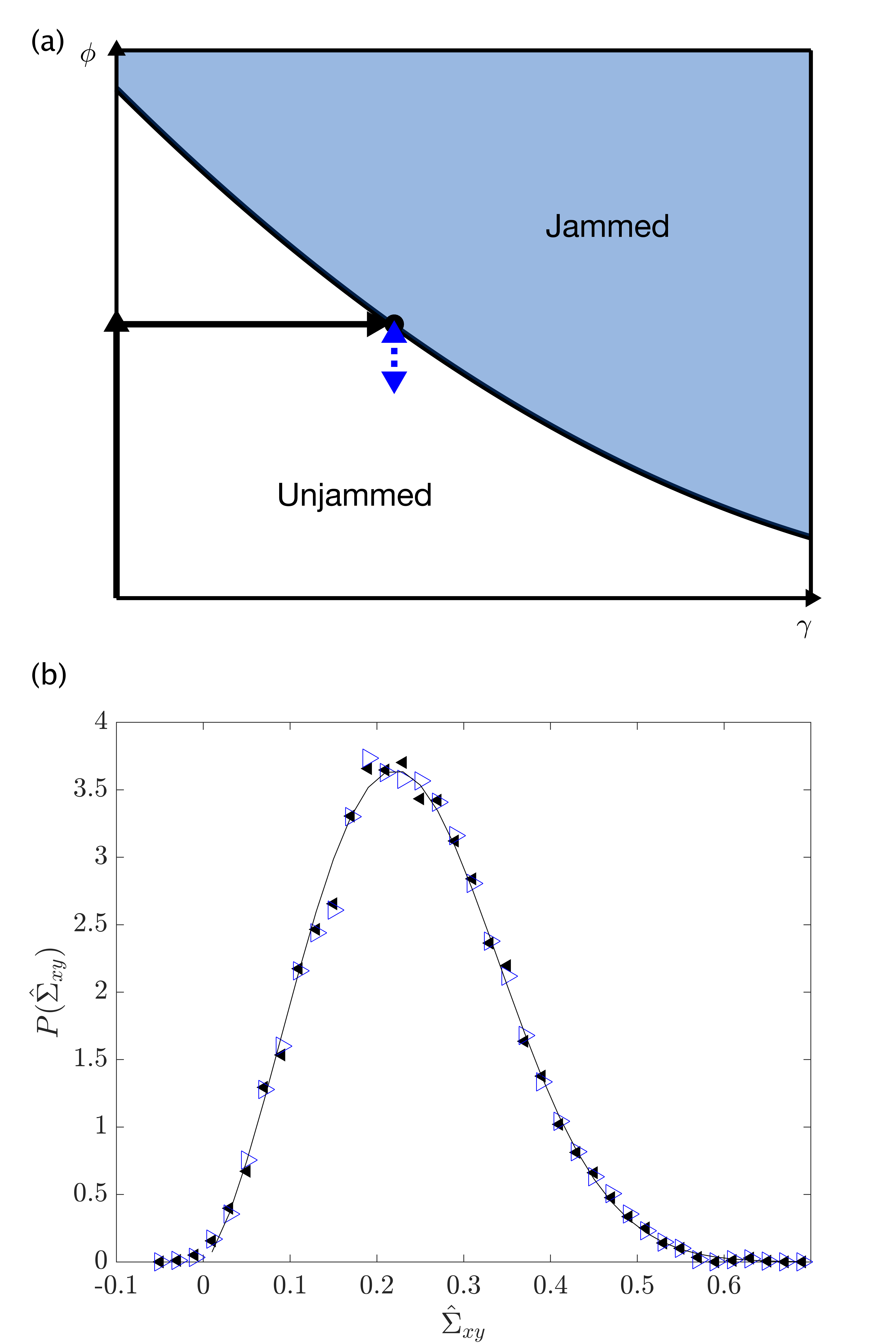}
}
\caption{(a) Illustration of protocol $A$ where we first generate 
a shear jammed packing (solid black lines), then decompress the shear jammed 
packing by $\Delta \phi$ and recompress it by $\Delta \phi$ to jamming onset
(blue dashed line). (b) Probability distribution of the shear stress 
anisotropy $P({\hat \Sigma}_{xy})$ for the original shear jammed packings 
(leftward filled triangles) and those generated using protocol $A$ (open 
rightward triangles) for systems with $N=16$ and $\mu=0.1$. The solid 
line is a Weibull distribution with scale and shape parameters $\lambda 
\sim 0.27$ and $k \sim 2.5$, respectively. 
}
\label{fig:6}       
\end{figure}

In Fig.~\ref{fig:6} (a), we illustrate protocol $A$. We decompress
each shear jammed packing at fixed $\gamma$ by $\Delta \phi \sim
10^{-8}$ that corresponds to the largest overlap, so that none of the
particles overlap and all of the tangential displacements are set to
zero. We then recompress each packing by $\Delta \phi$ in one step and
perform energy minimization. In Table~\ref{tab:1}, we show that out of
the original $8925$ shear jammed packings, protocol $A$ returned
$99\%$ compression rejammed packings with the same contact networks as
the original shear jammed packings and only $1\%$ of the compression
rejammed packings possessed different contact networks. None of the
packings were unjammed after applying protocol $A$. Even though the
memory of the mobility distribution of the original shear jammed
configuration is erased using protocol $A$, we show in
Fig.~\ref{fig:6} (b) that the distributions of the shear stress
anisotropy $P({\hat \Sigma}_{xy})$ are very similar for the original
shear jammed and compression rejammed packings. (We do not include the
small number of rejammed packings with different contact networks and
the unjammed packings in the distributions $P({\hat \Sigma}_{xy})$.)
In particular, both the compression rejammed packings and the original
shear jammed packings possess ${\hat \Sigma}_{xy} >0$, and thus the
distributions have nonzero means, $\langle {\hat \Sigma}_{xy} \rangle
> 0$.  This result implies that there is not a fundamental difference
between shear and compression jammed configurations, since the
isotropic compression protocol can generate ``shear jammed''
configurations.

\begin{figure}
\resizebox{1.0\hsize}{!}{%
  \includegraphics{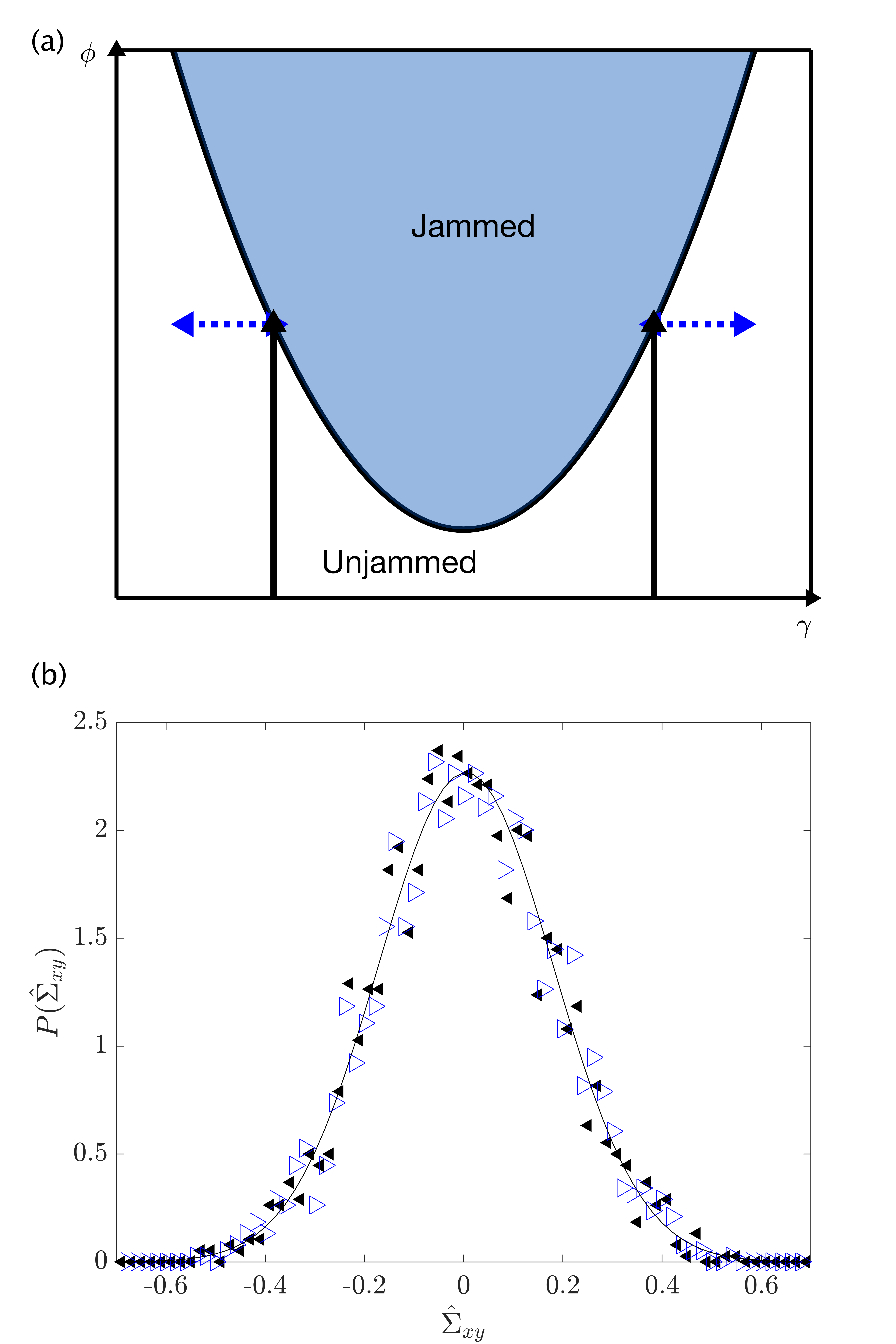}
}
\caption{(a) Illustration of protocol $B$ where we first generate 
compression jammed packigns (solid black lines). The compression jammed 
packings possess either $d\phi_J/d\gamma < 0$ (left) or 
$d\phi_J/d\gamma >0$ (right). For packings with $d\phi_J/d\gamma < 0$, 
we apply simple shear to the left by $\Delta \gamma$ to unjam them and 
then rejam them by applying $\Delta \gamma$ to the right (dashed blue lines
on the left). For packings with $d\phi_J/d\gamma > 0$, 
we apply simple shear to the right by $\Delta \gamma$ to unjam them and 
then rejam them by applying $\Delta \gamma$ to the left (dashed blue lines 
on the right). (b) Probability distribution of the shear stress 
anisotropy $P({\hat \Sigma}_{xy})$ for the original compression jammed 
packings 
(leftward filled triangles) and those generated using protocol $B$ (rightward 
open triangles) for systems with $N=16$ and $\mu=0.1$ The solid line 
is a Gaussian distribution with zero mean and standard deviation $\Delta \sim
0.2$.
}
\label{fig:7}       
\end{figure}

We now consider a related protocol where we shear unjam compression
jammed packings and then apply simple shear to rejam them.  In
Fig.~\ref{fig:7} (a), we illustrate protocol $B$. We first generate an
ensemble of compression jammed packings. Compression jammed packings
can jam on either side of the parabolic geometrical families
$\phi_J(\gamma)$; roughly half with $d\phi_J/ d\phi < 0$ and half with
$d\phi_J/ d\phi > 0$. For packings with $d\phi_J/ d\phi < 0$, we shear
by $\Delta \gamma \sim 10^{-8}$ in the negative strain direction to
unjam the packing.  For packings with $d\phi_J/ d\phi > 0$, we shear
by $\Delta \gamma \sim 10^{-8}$ in the positive strain direction to
unjam the packing. In both cases, to unjam the system, we apply simple
shear strain in extremely small increments $\delta \gamma = 10^{-12}$,
with each followed by energy minimization, until $U/N < U_{\rm tol}$.
After unjamming, we reset the tangential displacements at each nascent
contact to zero.  We then rejam the packings by applying the total 
accumulated shear strain $\Delta
\gamma$ in a single step in the opposite direction to the original
one, which allows the system to return to the same total strain, and
perform energy minimization. 

In Table~\ref{tab:1}, we show that out of the original $1987$
compression jammed packings, protocol $B$ returned $96\%$ shear
rejammed packings with the same contact networks as the original compression
jammed packings and only $4\%$ shear rejammed packings with
different contact networks. None of the packings generated using
protocol $B$ were unjammed. As shown in Fig.~\ref{fig:7} (b), the
distribution $P({\hat \Sigma}_{xy})$ of shear stress anisotropies is
nearly identical for the original jammed packings and the
rejammed packings. In both cases, $P({\hat \Sigma}_{xy})$ is
a Gaussian distribution with zero mean. This result emphasizes that isotropic
stress distributions can be generated using a shear jamming
protocol (when we consider shear jamming in both the positive and
negative strain directions).

\begin{figure}
\resizebox{1.0\hsize}{!}{%
  \includegraphics{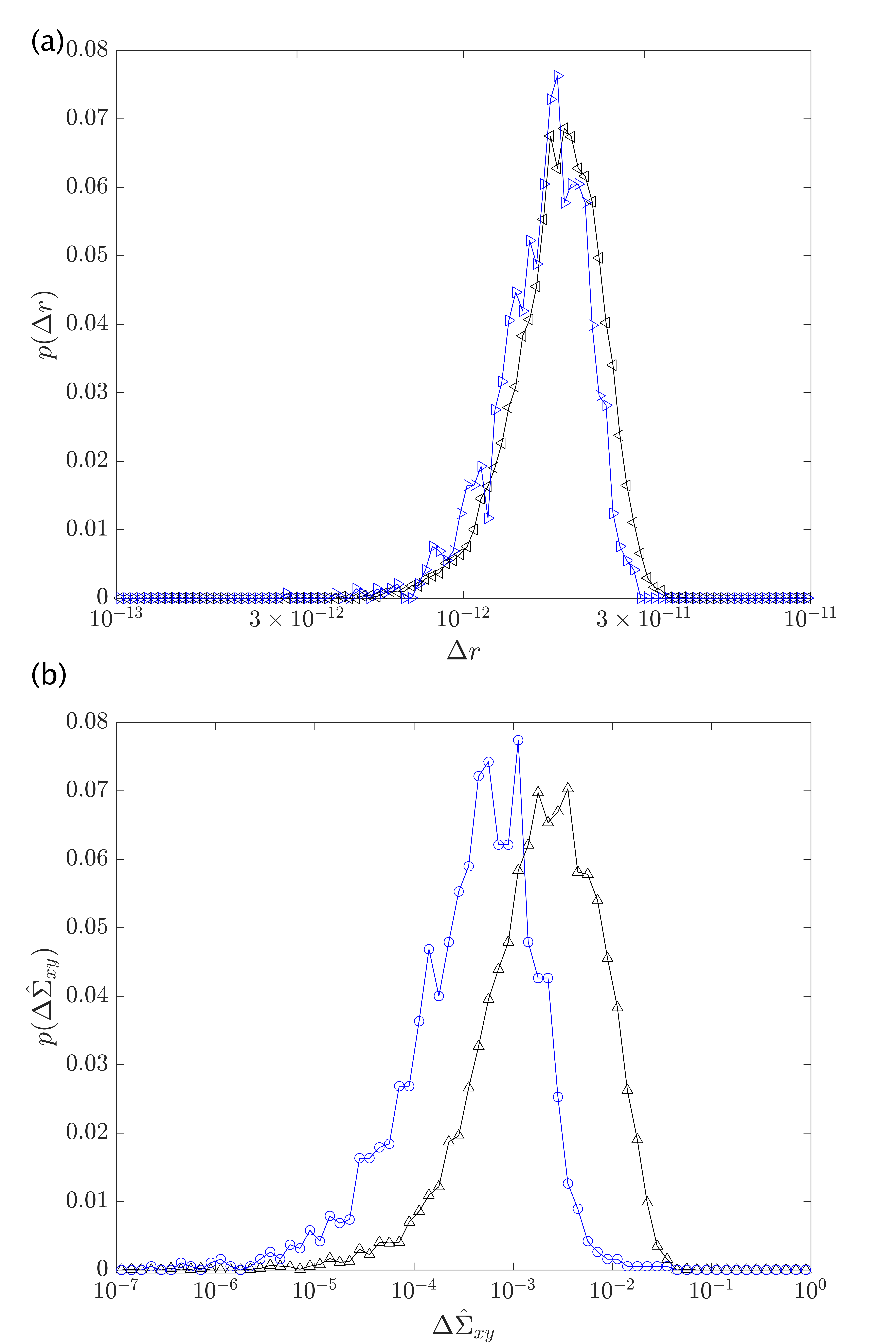}
}
\caption{(a) The frequency distribution $p(\Delta r)$ of the root-mean-square 
deviations in the positions
of the disks between shear jammed packings and those generated 
using 
protocol $A$ (triangles) and between compression jammed packings 
and those generated using protocol $B$ (circles). (b)
The frequency distribution $p(\Delta {\hat \Sigma}_{xy})$ of the 
root-mean-square 
deviations in the stress anisotropy between shear jammed packings and 
those generated using 
protocol $A$ (triangles) and between compression jammed packings 
and those generated using protocol $B$ (circles). For 
the data in both panels, $N=16$ and $\mu=0.1$. 
}
\label{fig:8}       
\end{figure}

We now compare directly the structural and mechanical properties of
the original shear jammed packings and those generated using protocol
$A$ and the original compression jammed packings and those generated
using protocol $B$. We calculate the root-mean-square deviation in
the particle positions, 
\begin{equation}
\label{rmsd}
\Delta r = \sqrt{N^{-1} \sum_{i=1}^N \left( {\vec
    r}_i^{A,B}- {\vec r}_i^{SJ,CJ} \right)^2},
\end{equation} 
and shear stress anisotropy, 
\begin{equation}
\label{rmsds}
\Delta {\hat \Sigma}_{xy} = \sqrt{ \left( {\hat
    \Sigma}_{xy}^{A,B} - {\hat \Sigma}_{xy}^{SJ,CJ} \right)^2}, 
\end{equation}
between the original shear jammed (SJ) packings and the packings
generated using protocol $A$ and the original compression jammed (CJ)
packings and the packings generated using protocol $B$. In
Fig.~\ref{fig:8} (a), we show the frequency distribution of the deviations in the particle positions
$\Delta r$ for systems with $N=16$ and $\mu=0.1$.  $\langle \Delta r \rangle \sim
2 \times 10^{-12}$ is extremely small, near numerical precision.  Thus, the
shear jammed packings and those generated via protocol $A$ have nearly
identical disk positions, and the compression jammed packings and
those generated via protocol $B$ have nearly identical disk positions.

We perform a similar comparison for the stress anisotropy (for systems
with $N=16$ and $\mu=0.1$) in Fig.~\ref{fig:8} (b). Even though the
disk positions are nearly identical between the shear jammed and
compression re-jammed packings, the typical root-mean-square
deviations in the stress anisotropy $\langle \Delta {\hat \Sigma}_{xy}
\rangle \sim 10^{-2.5}$ is finite. The stress anisotropy fluctuations
are nonzero because packings of frictional disks with the {\it same}
particle positions can have multiple solutions for the tangential
forces as shown using the force network ensemble~\cite{Ref28}. We find similar results for the differences in the stress
anisotropy between the compression jammed packings and the shear
re-jammed packings, however, the fluctuations are an order of
magnitude smaller with $\langle \Delta {\hat \Sigma}_{xy} \rangle \sim
10^{-3.5}$.  In contrast, when $\mu=0$, we find that $\langle \Delta {\hat
  \Sigma}_{xy} \rangle \sim 10^{-7}$ (nearly four orders of magnitude
smaller) for shear jammed packings and packings generated via protocol
$A$ with $\Delta r < 10^{-12}$.

\begin{figure}
\resizebox{1.0\hsize}{!}{%
  \includegraphics{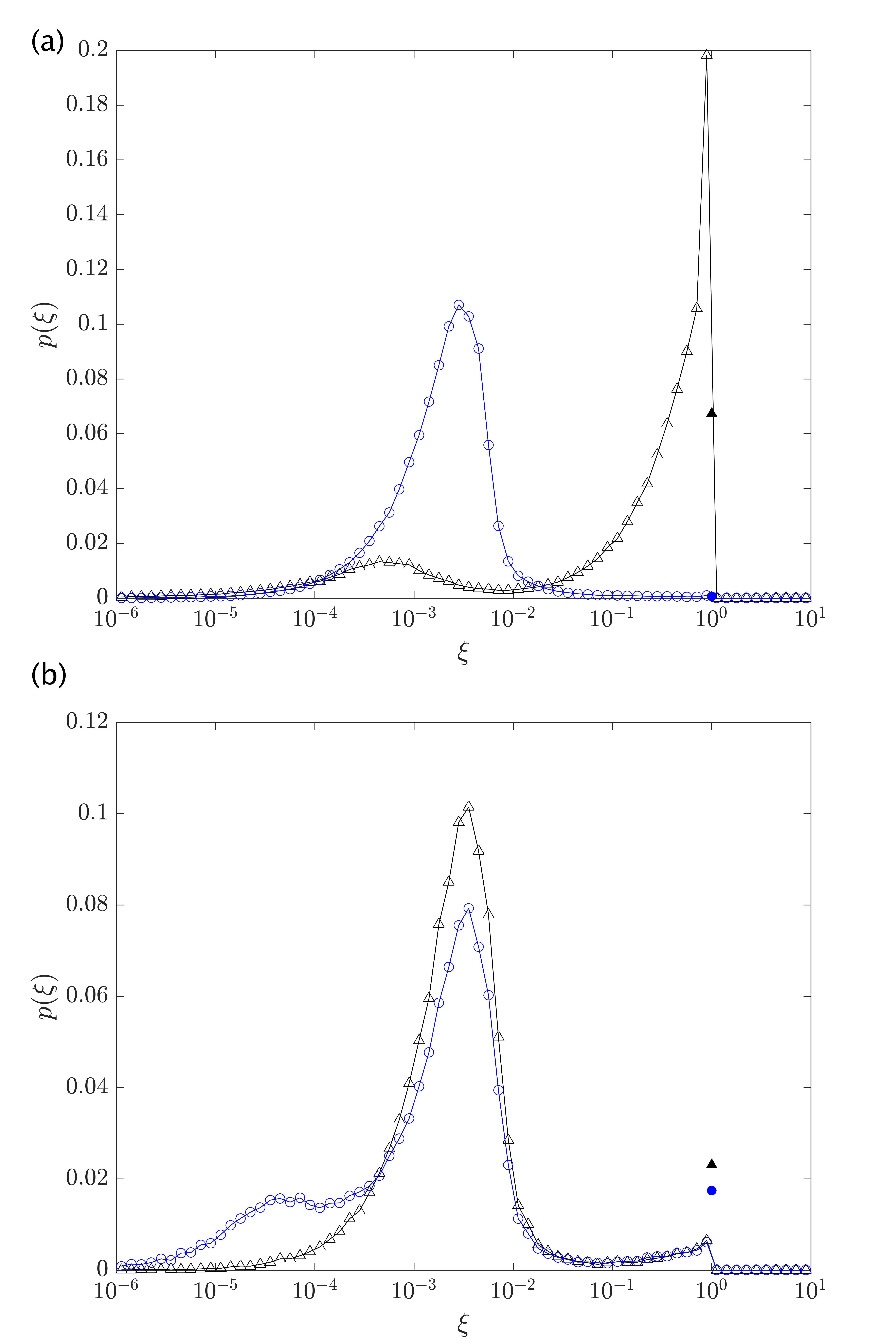}
}
\caption{The frequency distribution of the mobility 
$p(\xi)$, where $\xi=f^t_{ij}/\mu 
f^n_{ij}$ 
for each contact between disks $i$ and $j$, for shear jammed packings 
(open triangles) and compression re-jammed 
packings (open circles) with $N=16$ and $\mu=0.1$. (b) $p(\xi)$ 
for compression jammed packings (open triangles) and shear re-jammed 
packings (open circles) with $N=16$ and $\mu=0.1$. The filled symbols 
indicate the frequency of contacts that slid with $f^t_{ij} = \mu f^n_{ij}$.}
\label{fig:9}       
\end{figure}

We also compare the distributions of the mobility at each contact $\xi
= F^t_{ij}/\mu F^n_{ij}$ for the shear jammed packings and the
compression re-jammed packings, as well as the compression jammed
packings and the shear re-jammed packings. In Fig.~\ref{fig:9} (a), we
show that the original shear jammed packings have a significant number
of contacts that are near sliding with $\xi \sim 1$ and a smaller
fraction with $\xi \sim 10^{-3}$. However, the compression re-jammed
packings have essentially no sliding contacts, and instead most
contacts possess $\xi \sim 10^{-3}$. Thus, we find that the jamming
protocol can have a large effect on the contact mobility distribution.
Again, the abundance of tangential force solutions gives rise to the
shear stress anisotropy fluctuations even for packings with nearly
identical disk positions. In Fig.~\ref{fig:9} (b), we show $P(\xi)$
for the original compression jammed packings and the shear re-jammed
packings. These distributions are similar with a small fraction of
sliding contacts and abundance of contacts with $\xi \sim
10^{-3}$. This result is consistent with the fact that the stress
anisotropy fluctuations between compression jammed and shear re-jammed
packings are smaller compared to the stress anisotropy fluctuations
between shear jammed and compression re-jammed packings.

\begin{table}
\caption{(first row) Comparison of the contact networks (CN) for the original 
shear jammed (SJ) packings and compression rejammed packings. (second 
row) Comparison of the contact networks for the original compression 
jammed (CJ) packings and shear rejammed packings.}
\label{tab:1}       
\begin{tabular}{llll}
\\
\\
\hline\noalign{\smallskip}
SJ & same CN & different CN & Unjammed \\
\noalign{\smallskip}\hline\noalign{\smallskip}
8925 & 8875 & 50 & 0 \\
\noalign{\smallskip}\hline
\noalign{\smallskip}
CJ & same CN & different CN & Unjammed \\
\noalign{\smallskip}\hline\noalign{\smallskip}
1987 & 1899 & 88 & 0 \\
\noalign{\smallskip}\hline
\noalign{\smallskip}
\end{tabular}
\end{table}


\section{Conclusion and Future Directions}
\label{sec:4}

\begin{figure}
\resizebox{1.0\hsize}{!}{%
  \includegraphics{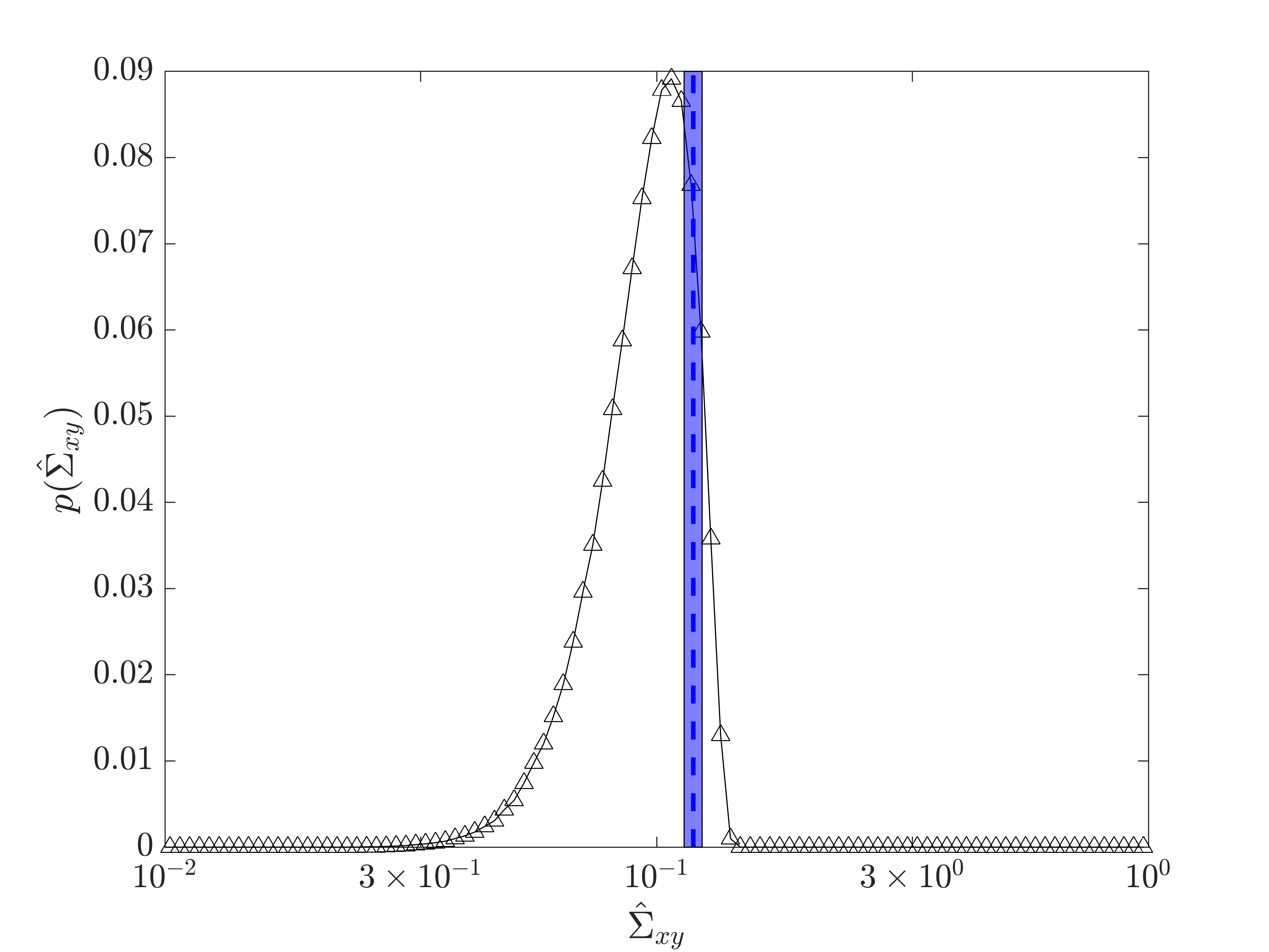}
}
\caption{The frequency distribution of the shear stress anisotropy 
$p(\hat{\Sigma}_{xy})$ calculated from the null space solutions
for a single compression jammed packing (open triangles). The 
vertical dashed line at ${\hat \Sigma}_{xy} \approx 0.12$ is the stress 
anisotropy of the given compression jammed packing and the shaded blue 
region (with width $5 \times 10^{-3}$) indicates the fluctuations in the 
stress anisotropy obtained
by comparing the compresssion jammed and shear rejammed packings from the 
DEM simulations.}
\label{fig:10}       
\end{figure}

\begin{figure}
\resizebox{1.0\hsize}{!}{%
  \includegraphics{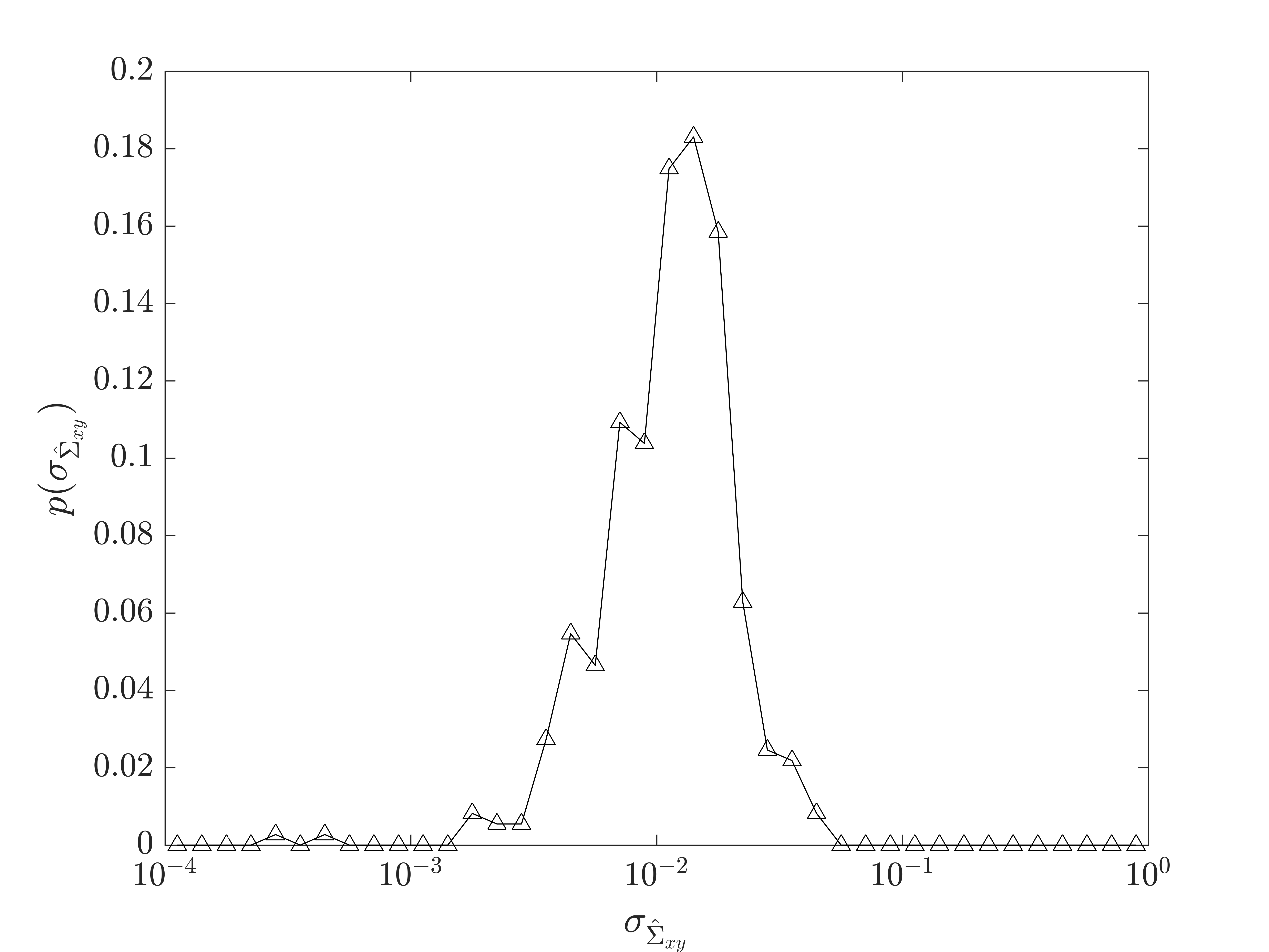}
}
\caption{The frequency distribution $p(\sigma_{{\hat \Sigma}_{xy}})$ of the 
standard deviation of the stress anisotropy from the null space solutions 
for each of the compression jammed packings. The peak in $p(\sigma_{{\hat \Sigma}_{xy}})$ is $\sigma_{{\hat \Sigma}_{xy}} \approx 10^{-2}$. 
}
\label{fig:11}       
\end{figure}

In this article, we used discrete element modeling simulations to
compare the structural and mechanical properties of jammed packings of
frictional disks generated via isotropic compression versus simple
shear.  We find that several macroscopic properties, such as the
average contact number $\langle z_J \rangle$ and packing fraction
$\langle \phi_J \rangle$ at jamming onset, are similar for both
packing-generation protocols. For both protocols, $\langle z_J(\mu)
\rangle$ varies from $4$ to $3$ in the low- and high-friction limits
with a crossover near $\mu_c \approx 0.1$. $\langle \phi_J(\mu) \rangle$
varies from $\sim 0.835$ to $0.76$ in the low- and high-friction
limits with a similar crossover value of $\mu_c$.

The average stress state of mechanically stable (MS) packings
generated via isotropic compression is different than that for MS
packings generated via simple shear. The average stress anisotropy
$\langle {\hat \Sigma}_{xy} \rangle > 0$ for MS packings generated via
shear, but $\langle {\hat \Sigma}_{xy} \rangle = 0$ for packings
generated via isotropic compression. Isotropic compression can sample
MS packings with both signs of ${\hat \Sigma}_{xy}$, whereas simple
shear (in one direction) samples packings with only one sign of the 
stress anisotropy.

To investigate in detail the differences in the stress state of MS
packings generated via simple shear and isotropic compression, we
developed two additional protocols.  For protocol $A$, we decompress
shear jammed packings so that the frictional contacts are removed and
then re-compress them to jamming onset. For protocol $B$, we shear
unjam MS packings generated via isotropic compression so that the
frictional contacts are removed, and then shear re-jam them. These
studies address an important question---to what extent can protocols $A$
and $B$ recover the contact networks and stress states of the
original jammed packings.  We find that even though protocols $A$ and
$B$ can recover the particle positions (and contact networks) of the
original jammed packings, the rejammed and original jammed packings
have small, but signficant differences in the stress anisotropy,
e.g. $\Delta {\hat \Sigma}_{xy} \sim 10^{-3.5}$-$10^{-2.5}$ for
systems with $\mu=0.1$.

To understand the stress fluctuations of frictional packings with
nearly identical particle positions, we carried out preliminary
studies of the null space solutions for force and torque balance on
all grains using the contact networks from the MS packings generated
via isotropic compression~\cite{Ref29}. For each packing of
frictional disks, force and torque balance on all grains can be
written as a matrix equation ${\cal A}_{lm} F_m = 0$, where ${\cal
  A}_{lm}$ is a $3N \times 2N_c$ constant matrix determined by the
contact network and $F_m$ is a $2N_c \times 1$ vector that stores the
to-be-determined normal and tangential force magnitudes $f^n_{ij}$ and
$f^t_{ij}$ at each contact. For frictional disk packings, the system
is underdetermined with $3N > 2N_c$.  Using a least-squares
optimization approach~\cite{Ref30}, we solve for the normal and tangential force
magnitudes such that $f^n_{ij} >0$, and $f^t_{ij} \leq
\mu f^n_{ij}$.

The stress anisotropy frequency distribution $p({\hat \Sigma}_{xy})$
from the null space solutions for an example compression jammed
packing (with $N=16$ and $\mu=0.1$) is shown in Fig.~\ref{fig:10}. We
find that the DEM-generated solutions belong to the set of null space
solutions, but there are many more. In particular, the width of
$p({\hat \Sigma}_{xy})$ is much larger than the width of the
distribution of the stress anisostropy obtained for the given
compression jammed packing from protocol $B$. We performed similar
calculations of the null space solutions for all compression jammed
packings. In Fig.~\ref{fig:11}, we show the frequency distribution of
the standard devivations $\sigma_{ {\hat \Sigma}_{xy}}$ of stress
anisotropy from the null space solutions over all of the compression
jammed packings. We find that the width of the fluctuations of the
stress anisotropy from the null space solutions for a given packing
are comparable to fluctuations of the stress anisotropy over all
compression jammed contact networks using DEM.  In future studies, we
will carry out similar calculations to understand how the fluctuations
in the stress anisotropy from the null space scale with system size
$N$ and friction coefficient $\mu$. For example, we will investigate
over what range of $N$ and $\mu$ are the null space stress aniostropy
fluctuations larger than the stress anisotropy fluctuations from
varying contact networks.  Addressing this question will allow us to
predict the differences in the structural and mechanical properties of
jammed packings of frictional particles that arise from the
packing-generation protocols, such as isotropic compression and simple
shear.

\begin{acknowledgments}
This research was sponsored by the Army
Research Laboratory under Grant No. W911NF-17-1-0164 (P.W., N.T.O.,
and C.S.O.).  The views and conclusions contained in this document are
those of the authors and should not be interpreted as representing the
official policies, either expressed or implied, of the Army Research
Laboratory or the U.S. Government.  The U.S. Government is authorized
to reproduce and distribute reprints for Government purposes
notwithstanding any copyright notation herein.  This work also
benefited from the facilities and staff of the Yale University Faculty
of Arts and Sciences High Performance Computing Center.  P.W. and
F.X. contributed equally to the paper.
\end{acknowledgments}


\begin{thebibliography}{}
%
%
\bibitem{Ref1}
O'Hern, C. S., Silbert, L. E., Liu, A. J., Nagel, S. R. Jamming at zero temperature and zero applied stress: The epitome of disorder. Physical Review E, \textbf{68}, 011306 (2003).
\bibitem{Ref2}
Makse, H. A., Johnson, D. L., Schwartz, L. M. Packing of compressible granular materials. Physical Review Letters \textbf{84}, 4160 (2000).
\bibitem{Ref3}
Bertrand, T., Behringer, R. P., Chakraborty, B., O'Hern, C. S., Shattuck, M. D. Protocol dependence of the jamming transition. Physical Review E \textbf{93}, 012901 (2016).
\bibitem{Ref4}
Bililign, E. S., Kollmer, J. E.,  Daniels, K. E. Protocol dependence and state variables in the force-moment ensemble. Physical Review Letters \textbf{122}, 038001 (2019).
\bibitem{Ref5}
Silbert, L. E. Jamming of frictional spheres and random loose packing. Soft Matter \textbf{6}, 2918-2924 (2010).
\bibitem{Ref6}
Miskin, M. Z., Jaeger, H. M. Evolving design rules for the inverse granular packing problem. Soft Matter \textbf{10}, 3708-3715 (2014).
\bibitem{Ref7}
Inagaki, S., Otsuki, M., Sasa, S. Protocol dependence of mechanical properties in granular systems. The European Physical Journal E \textbf{34}, 124 (2011).
\bibitem{Ref8}
Ciamarra, M. P., Nicodemi, M., Coniglio, A. Recent results on the jamming phase diagram. Soft Matter \textbf{6}, 2871-2874 (2010).
\bibitem{Ref9}
Majmudar, T. S., Behringer, R. P. Contact force measurements and stress-induced anisotropy in granular materials. Nature \textbf{435}, 1079 (2005).
\bibitem{Ref10}
Bi, D., Zhang, J., Chakraborty, B., Behringer, R. P. Jamming by shear. Nature \textbf{480}, 355 (2011).
\bibitem{Ref11}
Zhang, J., Majmudar, T., Behringer, R. Force chains in a two-dimensional granular pure shear experiment. Chaos: An Interdisciplinary Journal of Nonlinear Science \textbf{18}, 041107 (2008).
\bibitem{Ref12}
Zhang, J., Majmudar, T. S., Tordesillas, A., Behringer, R. P. Statistical properties of a 2D granular material subjected to cyclic shear. Granular Matter \textbf{12}, 159-172 (2010).
\bibitem{Ref13}
Kondic, L., Goullet, A., O'Hern, C. S., Kramar, M., Mischaikow, K., Behringer, R. P. Topology of force networks in compressed granular media. EPL (Europhysics Letters) \textbf{97}, 54001 (2012).
\bibitem{Ref14}
Gao, G.-J., Blawzdziewicz, J., O'Hern, C. S. Frequency distribution of mechanically stable disk packings. Physical Review E \textbf{74}, 061304 (2006).
\bibitem{Ref15}
Gao, G.-J., Blawzdziewicz, J., O'Hern, C. S. Geometrical families of mechanically stable granular packings. Physical Review E \textbf{80}, 061303 (2009).
\bibitem{Ref16}
Chen, S., Bertrand, T., Jin, W., Shattuck, M. D., O'Hern, C. S. Stress anisotropy in shear-jammed packings of frictionless disks. Physical Review E \textbf{98}, 042906 (2018).
\bibitem{Ref17}
Papanikolaou, S., O'Hern, C. S., Shattuck, M. D. Isostaticity at frictional jamming. Physical Review Letters \textbf{110}, 198002 (2013).
\bibitem{Ref18}
Song, C., Wang, P., Makse, H. A. A phase diagram for jammed matter. Nature \textbf{453}, 629 (2008).
\bibitem{Ref19}
Shen, T., Papanikolaou, S., O Hern, C. S., Shattuck, M. D. Statistics of frictional families. Physical Review Letters \textbf{113}, 128302 (2014).
\bibitem{Ref20}
Shaebani, M. R., Unger, T., Kert\'{e}sz, J. Extent of force indeterminacy in packings of frictional rigid disks. Physical Review E, \textbf{79}, 052302 (2009).
\bibitem{Ref21}
Handin, J. On the Coulomb Mohr failure criterion. Journal of Geophysical Research \textbf{74}, 5343-5348 (1969).
\bibitem{Ref22}
Cundall, P. A., Strack, O. D. A discrete numerical model for granular assemblies. Geotechnique \textbf{29}, 47-65(1979).
\bibitem{Ref23}
Xu, N., Blawzdziewicz, J., O Hern, C. S. Random close packing revisited: Ways to pack frictionless disks. Physical Review E \textbf{71}, 061306 (2005).
\bibitem{Ref24}
Silbert, L. E., Ertas, D., Grest, G. S., Halsey, T. C., Levine, D. Geometry of frictionless and frictional sphere packings. Physical Review E \textbf{65}, 031304 (2002).
\bibitem{Ref25}
Silbert, L. E., Ertas, D., Grest, G. S., Halsey, T. C., Levine, D., Plimpton, S. J. Granular flow down an inclined plane: Bagnold scaling and rheology. Physical Review E \textbf{64}, 051302 (2001).
\bibitem{Ref26}
Clark, A. H., Shattuck, M. D., Ouellette, N. T., O'Hern, C. S. Role of grain dynamics in determining the onset of sediment transport. Physical Review Fluids, \textbf{2}, 034305 (2017).
\bibitem{Ref27}
Peyneau, P. E., Roux, J. N. Frictionless bead packs have macroscopic friction, but no dilatancy. Physical Review E \textbf{78}, 011307 (2008).
\bibitem{Ref28}
Tighe, B. P., Snoeijer, J. H., Vlugt, T. J., van Hecke, M. The force network ensemble for granular packings. Soft Matter \textbf{6}, 2908-2917 (2010).
\bibitem{Ref29}
Vinutha, H. A., Sastry, S.  Force networks and jamming in shear-deformed sphere packings. Physical Review E \textbf{99}, 012123 (2019).
\bibitem{Ref30}
Coleman, T. F., Li, Y. A reflective Newton method for minimizing a quadratic function subject to bounds on some of the variables. SIAM Journal on Optimization \textbf{6}, 1040-1058 (1996).
\end{thebibliography}
\end{document}